\documentclass[fleqn,usenatbib]{mnras}

\usepackage{newtxtext,newtxmath}

\usepackage[T1]{fontenc}
\usepackage{ae,aecompl}

\usepackage{graphicx}	
\usepackage{amsmath}	

\usepackage{amssymb}	

\usepackage{hyperref}

\title[Karma asteroid family]{Analysis of the Karma asteroid family}

\author[D. Pavela et al.]{
Debora Pavela,$^{1}$\thanks{E-mail: debbiepavela2@gmail.com (DP)}
Bojan Novakovi\' c,$^{1}$ Valerio Carruba,$^{2}$ \& Viktor Radovi\'c$^{1}$
\\
$^{1}$Department of Astronomy, Faculty of Mathematics, University of Belgrade, Studentski trg 16, 
11000 Belgrade, Serbia\\
$^{2}$Sao Paulo State University (UNESP), School of Natural Sciences and Engineering, Guaratingueta, SP, 12516-410, Brazil\\
}

\date{Accepted 2020 November 19. Received 2020 November 17; in original form 2020 July 25}

\pubyear{2020}

\begin{document}
\label{firstpage}
\pagerange{\pageref{firstpage}--\pageref{lastpage}}

\maketitle

\begin{abstract}

The Karma asteroid family is a group of primitive asteroids in the middle part of the main belt, just at the outer edge of the 3J:1A mean-motion resonance. We obtained the list of the family members with 317 asteroids and estimated that it was formed by the catastrophic disruption of a parent body that was between 34 and 41~km in diameter. Based on the V-shape method, age of the Karma family is estimated to be about 137~Myr.
A detailed dynamical map of the region combined with numerical simulations allowed us to reconstruct the long-term dynamical evolution of the family, and to identify the mechanisms responsible for this evolution.
The numerical simulations successfully reproduced the main features in the orbital distribution of the family members but also showed that some regions of the Karma family could be missing. A more detailed analysis revealed that these regions likely consist of very dark objects, fainter than absolute magnitude $H$=17, that have not yet been detected. Based on the obtained results, we concluded that magnitude-frequency distribution of family members up to $H$=16~mag is neither affected by dynamical erosion nor observational incompleteness and therefore represents the result of collisional grinding of the original family population. Finally, we found that the Karma family have been supplying some asteroids to the near-Earth region via the 3J:1A resonance. Currently there should about 10 family members larger than 1~km in diameter, orbiting in the near-Earth space.

\end{abstract}




\section{Introduction}

An asteroid family is a group of asteroids with similar orbital elements, mostly formed by asteroid collisions. When the collision happens, fragments from the parent body are ejected in the nearby space, forming a cloud-like formation around the location where the collision took place. Such groups may be identified as asteroid families, in some cases even a billion years after the original collision. 

Studies of asteroid families are important from many points of view. For instance, identification of the families and their formation rate give an overview of how the collisional evolution affected the main asteroid belt \citep{bottke2015asteroidsIV}. Families allow studying impact physics involving the disruption of large bodies, and their ejected fragments, which can not be done in laboratories in the same size range. Also, spectroscopic observation of a family members can give us insights into the internal mineralogical structure of the parent body \citep{md2005a,masiero2015}. 

Families of primitive taxonomic composition are those formed by break-ups of dark carbonaceous parent asteroids. Primitive families located in the inner part of the main belt have received lots of attention in the recent years \citep[see][and references therein]{Morate2019}, mostly because they were identified as potential sources of two near-Earth asteroids (NEAs) (101955)~Bennu and (162173)~Ryugu, targets of the sample-return missions OSIRIS-REx and Hayabusa2, respectively \citep{2015Icar..247..191B}. Also, over the last decade, increased attention has been paid to the primitive families located in the outer part of the main belt \citep{DePra2020}, mainly due to their possible connection with main-belt comets \citep[see e.g.][]{Hsieh2018}, and more recently with Jupiter family comets \citep{Hsieh2020}. 

Here we study the (3811)~Karma asteroid family, a group of primitive asteroids situated in the middle part of the main asteroid belt. This is an interesting family, that is important to study for many reasons. Families interacting with week and moderately strong secular resonances, such as the Karma, may in some cases help us to reconstruct parameters of the collision event that produced the family \citep{carruba2018}, and to improve our knowledge of the impact physics. Also, being located close to the edge of the powerful 3J:1A mean motion resonance with Jupiter, the Karma family likely supplied some asteroids to the near-Earth region. Moreover, observational studies of the Karma family asteroids may provide important clues about the innermost icy-asteroids in the main belt.
Still, the Karma family has not attracted much attention among researchers in the past, probably due to a limited amount of data available about the asteroids belonging to it. This situation has changed in the meantime, and the data available now allow to study the family from different perspectives, and to draw reliable conclusions.
In this work, we present a comprehensive study of the Karma family, focusing on its dynamical evolution.

\section{Karma family: basic properties}

\subsection{Membership and spectral reflectance characteristics}
\label{ss:members}

Any analysis of an asteroid family depends on the accuracy of identification of its members. Detection of asteroid families and identification of their members is typically performed using proper orbital elements: semi-major axis $a_p$, eccentricity $e_p$, and inclination $i_p$ \citep{Zappala1990,milani2014,nes2015}, though in principle this could be done also in the space of proper frequencies \citep{2007A&A...475.1145C}. Here we used the catalogue of the proper elements of numbered and multi-opposition asteroids, with 631,226 objects in total, available at the Asteroid Families Portal.\footnote{\url{http://asteroids.matf.bg.ac.rs/fam/properelements.php}}

\begin{figure}
	\includegraphics[width=0.95\columnwidth, angle=0]{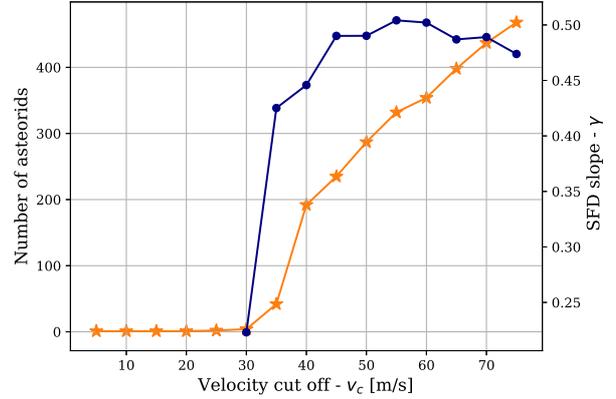}
    \caption{Number of asteroids associated with the Karma family (shown as orange stars), and a value of the slope $\gamma$ of the cumulative magnitude distribution fitted in the range $H\in[15, 17]$ (presented as dark blue circles), as a function of the HCM's cut-off velocity $v_c$. Adopted nominal cut-off is $v_{c}=55$~$ms^{-1}$.}
    \label{fig:nfv}
\end{figure}

The most widely used algorithm to identify members of an asteroid family is the Hierarchical clustering method \citep{Zappala1990}.
In this method, distances in terms of velocity among all combinations of asteroids are computed, and objects mutually closer than a given threshold distance $v_{c}$ (often called the cut-off velocity) are assumed to be family members. Adopting an appropriate $v_{c}$ is, however, not always straightforward. Usually, a plot showing how the number of family members $N$ depends on $v_{c}$ is created, and the nominal $v_{c}$ is adopted as a centre of the interval where $N$ grows only moderately \citep[e.g.][]{nov2011}. However, in the case of the Karma family, such a \textit{plateau} does not exist. As can be seen in Fig.~\ref{fig:nfv}, a core of the family is identified at $35$~$ms^{-1}$, and the number of associated members grows steadily until 80~$ms^{-1}$, when the family merges with a local background population of asteroids. Therefore, in this case any value between $40$ and $75$~$ms^{-1}$ is a possible nominal cut-off distance $v_{c}$. Here we adopt a value of $v_{c}=55$~$ms^{-1}$ as the nominal cut-off, which connects 332 asteroids to the Karma family.\footnote{Note that at this cut-off the lowest numbered asteroid associated with the family is (500)~Selinur. This object is however definitely an interloper as its spectral type \citep{Lazzaro2006} and albedo \citep{wise2011} are both incompatible with C-type taxonomic classification of the Karma family.} This is a relatively conservative value that allows reducing the number of interlopers associated with the family, with a cost of potentially missing some family members.

\begin{figure}
	\includegraphics[width=0.7\columnwidth, angle=-90]{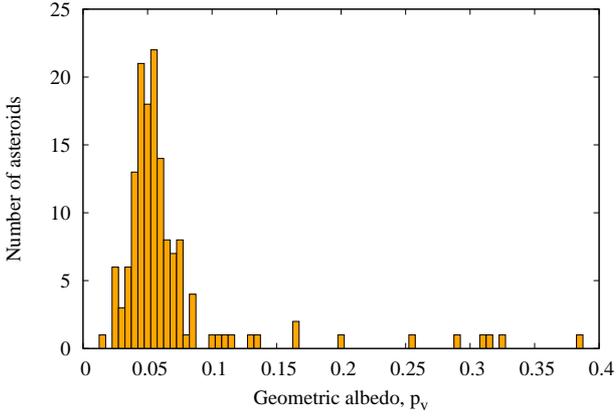}
    \caption{The number frequency distribution of albedos of the Karma family members. The family belongs to C taxonomic type, with members' geometric visual albedos smaller than 0.1. The objects with larger albedos are most likely to be interlopers.}
    \label{fig:albedo}
\end{figure}

However, even when an asteroid family is determined in the proper element space, not all associated asteroids are real family members. Despite using a somewhat lower value of the cut-off velocity $v_{c}=55$~$ms^{-1}$, the list of the family members obtained by the HCM unavoidably includes some interlopers. As members of an asteroid family are typically homogeneous in composition \citep{md2005b,parker2008,2013MNRAS.433.2075C}, to find interlopers of the family we have to consider information on the composition of each member \citep[e.g.][]{masiero2015,radovic2017}. Surface reflectance characteristics of family members have been used to improve the membership, by identifying and removing potential interlopers.The most common information in this respect is asteroid albedo. An average albedo of family members can tell us to which spectral complex family belongs.
Comparing the average family albedo and measured albedos of individual family members, potential interlopers can be identified.\footnote{Due to the limitations of asteroid surveys, albedo is not measured for all asteroids. This can result in having more interlopers in the family than initially identified. Apart from this, the efficiency of the identification of interlopers depends on how different in composition family members are concerning nearby non-family members. As in many cases, families are taxonomically indistinguishable from non-family asteroids \citep{Erasmus2019}, only in cases when family and background asteroids are of different composition, interlopers could be efficiently identified.}
In particular, to this purpose, we have used WISE albedos \citep{wise2011} and found measured albedos for 146 family members. Because Karma family consists of dark asteroids, as can be seen in Fig.~\ref{fig:albedo}, we removed from the family 15 asteroids with an albedo above 0.1. 
After removing these objects, we found the mean geometric albedo of the family of $p_v=0.055\pm0.014$. Also, we have investigated available colours of family members in the 4th Release of the Sloan Digital Sky Survey (SDSS) Moving Object Catalog \citep{sdss}. We found that this catalogue contains colours of 55 family members. Based on this data, and following \citet{radovic2017}, we were able to reliably identify only one potential interloper, the asteroid (38685) 2000~QP$_{9}$. However, as this object was already marked as the interloper based on its albedo, no additional asteroid has been removed from the family based on the SDSS colours.

\begin{figure}
	\includegraphics[width=0.7\columnwidth, angle=-90]{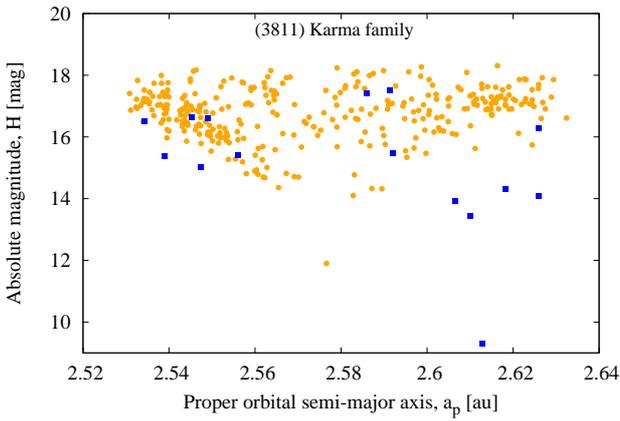}
    \caption{Karma family members determined by the HCM with $v_{c}=55$~$ms^{-1}$, projected onto a plane of proper semi-major axis versus absolute magnitude. The family members are shown as orange circles, while the identified interlopers are marked as blue squares.} 
    \label{fig:aH}
\end{figure}

In Fig.~\ref{fig:aH} we show all asteroids associated to the family in the
semi-major axis versus absolute magnitude plane. Due to the Yarkovsky effect, an asteroid family in this plane\footnote{Except in the semi-major axis versus absolute magnitude plane, the V-shape structures are also visible in the semi-major axis versus inverse diameter plane. In the latter case, the boundaries of the V-shaped region are straight lines, because the Yarkovsky induced drift rate scales inversely with diameter.} typically form a V-shape like structure, and real family members are expected to be located inside the boundaries of the
V-shape. We note that most of the identified interlopers are located outside
the V-shape boundaries, providing an additional evidence that these objects
are not real family members.

\begin{figure}
	\includegraphics[width=1\columnwidth, height=9cm]{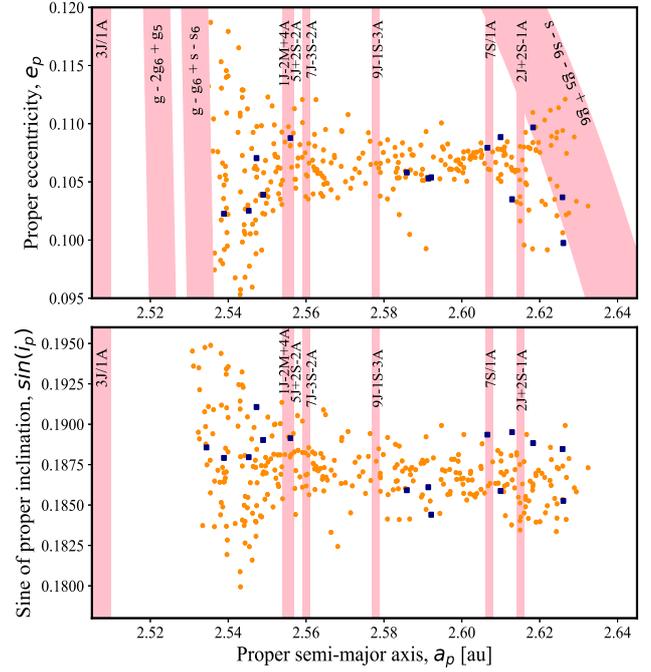}
    \caption{Karma family members determined by the HCM with $v_{c}=55$~$ms^{-1}$. Projection is onto a plane of the proper semi-major axis and sine of proper inclination (top panel) and a plane of the proper semi-major axis and proper eccentricity (bottom panel). The family members are shown as orange circles, while the identified interlopers are marked as blue squares. In addition, locations of major resonances are labelled and shown on the plot. Note that the locations of the secular resonances are labelled only in the upper panel.}
    \label{fig:aei}
\end{figure}

A simple analysis of the distribution of family members in the $a_p-e_p$ and $a_p-\sin(i_p)$ planes, shown in Fig.~\ref{fig:aei}, reveals some interesting features. Family members are more dispersed at family's edges, in terms of the proper semi-major axis. For instance, in the $a_p-e_p$ plane (top panel in Fig.~\ref{fig:aei}) spreading around $2.54$~au is about twice as large as the one at about $2.58$~au. Although less noticeable, a similar pattern is also visible near the outer edge of the family, around $a_p = 2.62$~au. Likewise, analogous features are visible in the $a_p-\sin(i_p)$ plane. All these point out towards significant dynamical evolution of the family.

\begin{figure}
	\includegraphics[width=0.7\columnwidth, angle=-90]{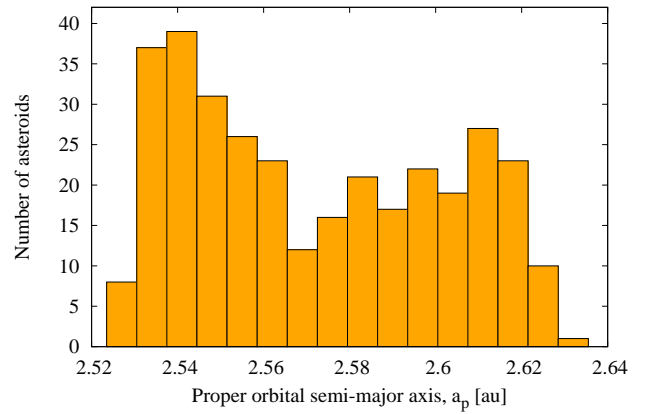}
    \caption{Histogram of the proper semi-major axis of the Karma family members.}
    \label{fig:an}
\end{figure}

The number-frequency distribution of family members in terms of the semi-major axis, shown in Fig.~\ref{fig:an}, reveals a concentration of asteroids near the family edges. The fact that small members accumulate at the extreme semi-major axis values has been already noted in some families \citep[see e.g.][and references therein]{vok2015}. This formation of "ears" is a consequence of the Yarkovsky-O'Keefe-Radzievskii-Paddack (YORP) effect, which drives asteroids' obliquity toward $0$ or $180$ degrees (spin axis perpendicular to the orbital plane), and therefore maximizes the Yarkovsky drift \citep{Vokrouhlicky_2006b}.

The distribution in Fig.~\ref{fig:an} is also slightly asymmetric with respect to the centre of the family, with the left side of the family in term of the proper semi-major axis being overpopulated by about 15 per cent. A similar pattern has been also found in other families and the most recent in the Clarissa family \citep{2020AJ....160..127L}.
This might be a consequence of dynamical erosion of the family, but it might also be related to an impact geometry during the collisional event that produced more retrograde than prograde rotators \citep[see e.g. discussion in][]{milani2014,2020AJ....160..127L}.

\subsection{Magnitude distribution and parent body size}
\label{ss:mfd_pb}

Important information to gain from each asteroid family is its magnitude (size) distribution \citep{masiero2015}. Fig.~\ref{fig:nfv} shows how a slope $\gamma$ of the cumulative magnitude-frequency distribution for $H\in [15,17]$, changes for different values of the cut-off velocity used to define the family. For our nominal family, defined at $v_c$ = 55~m$s^{-1}$, the slope is $\gamma=0.50$. This relatively shallow slope is not expected for <200 Myr old families like the Karma family (see Section~\ref{ss:age}), and it may suggest that some small members of the family are missing. 

To examine this possibility, we identify the absolute magnitude completeness limit of the family, using an approach similar to the one proposed by \citet{nov2010}. We found that this limit, in the range of the semi-major axis covered by the family, is at $H$=16.1~mag. Since the orbital inclination of the family is around 11 degrees, and most sky surveys for asteroids are in favour of detecting asteroids moving close to the ecliptic plane, the completeness limit of the Karma family should be slightly lower, around an absolute magnitude of $H$=16.  Therefore, we expect some of the family members to be missing in the magnitude range 15-17.
A possible additional loss of the family members due to their dynamical evolution will be further analysed in Section~\ref{s:num_sim}. 

The size of the parent body can be estimated in at least three different ways:
(i) as a sum of diameters of two largest fragments \citep{tanga1999}, (ii) from SPH simulations \citep{durda2007}, and (iii) simply by summing up volumes of all family members and by determining a diameter of resulting spherical body \citep{nov2010}. For simplicity, here we use the approaches (i) and (iii) and found that the Karma family parent body was between 34 and 41~km in diameter. Though the first method that we used here generally underestimates the size of the parent body due to the members not associated to the family, the second one should not be affected as long as the two largest family members are reliably identified. Nevertheless, we note that, in principle, the parent body might be a few kilometres in the diameter larger. 

\subsection{Dynamical environment}
\label{ss:dyn_env}

To study the physics of the collisional events from which families originate, the initial state of these families need to be reconstructed. In this respect, both the collisional and dynamical evolution have to be considered. Dynamical evolution is caused by different gravitational and non-gravitational perturbations resulting in modification of an initial orbital structure of a family over time. The relevant gravitational effects are close encounters and resonances, while the main non-gravitational force is the Yarkovsky effect. The resonances mainly change orbital inclination and eccentricity, meanwhile, the Yarkovsky effect results in the dispersion of family members in terms of orbital semi-major axis \citep{2006AREPS..34..157B}. On the other hand, effects of close encounters may affect all three proper orbital elements, but they depend on the mass of a perturbing body and generally are a less important type of the perturbations in the main belt. These mechanisms may cause significant orbital evolution of asteroid families, which in turn may result in unusual shapes \citep{bottke2001,nov2015ApJ}, and/or removal of a significant fraction of family members, that is in some cases connected with a supply of meteorites to the Earth \citep{zappala1998,vok2017,2019SciA....5.4184S}. The above-described evolution could be successfully modelled, only if the ages of the asteroid families are known. Today's available methods for age determination can provide ages for practically all families \citep[see e.g.][]{nes2015,spoto2015,milani2017}. However, some of these estimations have large uncertainties, which can affect the model of family evolution. Therefore, of particular interest are families whose ages could be reasonably well constrained.

Level of the dynamical evolution of an asteroid family depends on its age, but also on the dynamical characteristics of the region where the family is located. We identify various mean-motion resonances (MMRs) and some secular resonances (SRs) that cross the region. The most important resonances are shown in Fig.~\ref{fig:aei}. These include three-body MMRs \citep{nesmor1998}, between either Jupiter, Saturn and asteroid (2J +2S -1A, 9J -1S -3A, 7J -3S -2A, 5J +2S -2A) or Jupiter, Mars and asteroid (1J -2M +4A), and two body resonance with Saturn (7A:1S). The most important secular resonances\footnote{The $g, s, g_6, s_6, g_v, s_v$ denote proper frequencies of longitude of the ascending node and longitude of perihelion, of an asteroid, Saturn and Vesta, respectively.} are the $g-g_{6}+s-s_{6}$ and $s-s_6-g_5+g_6$ that intersects the family near the inner and outer edge, respectively (see Fig.~\ref{fig:aei}). Furthermore, the secular resonance $g-g_v+s-s_v$ with asteroid (4)~Vesta \citep{tn2016} is also crossing the family, but its importance seems to be minor.

The dispersion of the Karma family members close to the inner edge of the family is mostly caused by the 3J:1A resonance with Jupiter, and the $g-g_{6}+s-s_{6}$ secular resonance, while the main actor in scattering the family members close to the outer edge is probably the $s-s_6-g_5+g_6$ secular resonance.

\begin{figure}
	\includegraphics[width=0.7\columnwidth, angle=-90]{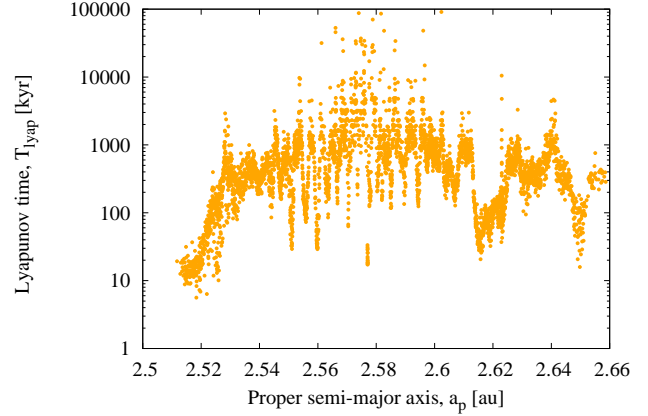}
    \caption{Lyapunov times in the region inhabited by the Karma family. The smaller values indicate less stable orbits.}
    \label{fig:lyapunov}
\end{figure}

In Fig.~\ref{fig:lyapunov} we show Lyapunov times\footnote{The Lyapunov time indicates the characteristic timescale on which a dynamical system becomes chaotic. Time limits on the predictability of orbital motion of an asteroid are typically several Lyapunov times.} as a function of the proper semi-major axis. These results suggest that most of the asteroids in this region reside on relatively stable orbits ($T_{lyap}>100$~kyr), but there is also a significant fraction of highly unstable orbits ($T_{lyap}<20$~kyr). The most chaotic orbits are associated with the 3J:1A resonance with Jupiter (($T_{lyap}\sim$10~kyr)

\subsection{Age}
\label{ss:age}

To reconstruct the dynamical evolution of the Karma family, the age of the family has to be known, however, we did not find any age estimation available in the literature. Therefore, for the family age estimation, we utilised the so-called V-shape method as proposed by \citet{spoto2015}. 

\begin{figure}
	\includegraphics[width=0.99\columnwidth, angle=0]{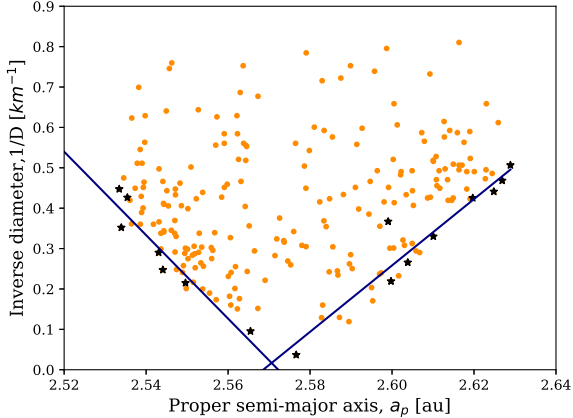}
    \caption{The V-shape of the Karma family. Family members are represented in orange, while black stars show family members of every bin that are used for fitting. }
    \label{fig:a1D}
\end{figure}

We first analyse the V-plot of the family looking for possible asteroids which are significantly displaced concerning the borders of the V-shape plot. However, apart from the interlopers already removed in Section~\ref{ss:members}, we did not find any additional outlier. The next step was to select the fitting range in terms of 1/D and to divide it in an appropriate number of bins. We found that boundaries of the Karma family V-shape are well defined for $1/D <$ 0.5~ms$^{-1}$, i.e. for objects larger than 2~km in diameter (Fig.~\ref{fig:a1D}). Accordingly, we divide this range of the 1/D axis into eight bins, in such a way that each bin contains roughly the same number of members. For the left side of the family, we select the minimum value of $a_p$ and the corresponding 1/D in each bin, while for the right side we select the maximum value of the proper semi-major axis and its corresponding value of 1/D. Then, we use the least-squares method to fit the data on both sides of the V-shape, with straight lines (see Fig.~\ref{fig:a1D}). This way, we obtain the slopes of the V-shape.

To compute the age of the family, we also need to consider the Yarkovsky effect. The estimated maximum value of the Yarkovsky induced secular drift $(da/dt)_{max}$ for a hypothetical family member of $D=1$~km in diameter is obtained using an available model of the Yarkovsky effect \citep{vok2015}. The adopted thermal parameters are $\rho_s = \rho_b =$1190~kg~m$^{-3}$ for the surface and bulk densities respectively, and thermal inertia of $\Gamma =$ 350~J~m$^{-2}$~s$^{-1/2}$~K$^{-1}$, based on the recent findings of the OSIRIS-REx and Hayabusa2 missions \citep{bennu2019,ryugu2019}. The resulting maximum drift speed of $1$~km asteroid is estimated to be about $8\times$10$^{-4}$~au~Myr$^{-1}$.

Combining the maximum Yarkovsky drift speed and the slopes of the V-shape plot, we estimate that the Karma family is $137\pm21$~Myr old. We will consider this value as a nominal age of the Karma family.

It should be noted, however, that any method for age determination has some limitations. The Yarkovsky-based V-shape methods depend on the Yarkovsky calibration, and, to some degree, on the initial state of the family. In the case of the Karma family, using the Yarkovsky calibration adopted by \citet{spoto2015}, of $(da/dt)_{max}$ about 6$\times$10$^{-4}$~au~Myr$^{-1}$, would result in an older age of $181\pm27$~Myr. On the other hand, the results obtained using the version of the method as proposed by \citet{spoto2015} do not directly depend on the initial size of a family, but the ages could still be somewhat overestimated in case the ejection velocity field produces an initial V-shape of the family, not related to the Yarkovsky effect. For these reasons, the real uncertainty of our age estimation might be a somewhat larger.

\section{Numerical simulations}
\label{s:num_sim}

\subsection{Method} 

The dynamical evolution is simulated by performing a set of numerical integrations, using the multi-purpose \textit{OrbFit} software package (available from http://adams.dm.unipi.it/orbfit/). The main dynamical model includes the gravitational effects of the Sun and seven major planets, from Venus to Neptune, plus the Yarkovsky non-gravitational effect. The Yarkovsky effect is implemented simply as a pure along-track acceleration, producing on average the same semi-major axis drift as expected from the Yarkovsky calibration. 

\subsection{Synthetic Karma family: Initial conditions}
\label{ss:initial_cond}

The size-frequency distribution of the test particles is produced from the magnitude-frequency distribution as it follows. We assume that the slope $\gamma = 0.50$ of the magnitude-frequency distribution obtained in Section~\ref{ss:mfd_pb} for $H\in[15,17]$~mag is valid over an extended range of absolute magnitudes, that is up to $H$=18.1~mag, what is approximately magnitude of the faintest known family member. Then, up to $H$ = 17~mag.  we took the absolute magnitudes of real family members, while based on the obtained slope of the distribution, we estimated that there are 554 family members in $H\in[17,18.1]$~mag range. Therefore, with 166 real family members identified for $H < $17~mag, this leads a total of 720 expected family members for $H \leqslant$ 18.1~mag. Finally, diameters of test particles representing the synthetic Karma family are obtained according to the following formula:
\begin{equation}
        D [km] = 1329 \frac{10^{-\frac{H}{5}}}{\sqrt{p_v}},
    \end{equation}
and using the mean geometric albedo of the family of $p_v=0.055$.

For collisionally formed asteroid families, the largest velocities at infinity $v_{\infty}$ of km-sized members with respect of the parent body, are empirically found to be roughly similar to the escape velocity from the parent body\footnote{We recall here that velocities at infinities with respect of the parent body of an asteroid family are sometimes called velocity change \citep[e.g.][]{nov2012}, or terminal ejection velocities \citep[e.g.][]{carruba2016}} \citep{nes2015,carruba2016}. The estimated size of the Karma family parent body is between 34 and 41~km (see Section~\ref{ss:mfd_pb}), and when combined with a density of 1190~kgm$^{-3}$, it yields an escape velocity approximately between 14 and 17~m$s^{-1}$. Based on this, we adopted a relatively compact initial family, defined by the ejection velocity parameter $V_{EJ}=$10~ms$^{-1}$ (see Eq.~\ref{eq:velocity}).

Knowing the ejection velocity parameter, and assuming that the initial ejection velocity field was isotropic, Gauss equations could be used to determine an ellipsoid in 3-D orbital space that represents the distribution of family members immediately after the fragmentation event. We have also assumed that the catastrophic disruption of the Karma family parent body occurred at the barycenter of the family, that is found to be near the present-day position of the asteroid (3811)~Karma, but does not exactly coincide with it.  

The studies of the size-velocity relationship for members of asteroid families \citep{cellino1999,carruba2016,bolin2018}, suggest that initial dispersion with respect to the centre of the family is inversely proportional to fragments' diameters \citep[see also][]{Marzari1995,vok2006,rospla2018}. To account for this, we defined the maximum velocities at infinity that a test particle can achieve as: 
 \begin{equation}
        V_{max} = V_{EJ} \left (D_{0}/D \right),
    \label{eq:velocity}
    \end{equation}
where $D$ is the estimated size of the body, and $D_0$ and $V_{EJ}$ are reference values, adopted here to be 2~km and 10~ms$^{-1}$, respectively.
Then, to each test particle representing the synthetic Karma family, we assign a random velocity at infinity between zero and its corresponding $V_{max}$.

In addition to the size and orbit distribution, to complete our initial conditions we need to assign a corresponding semi-major axis drift speed induced by the Yarkovsky effect, to each test particle. For this purpose, we used an isotropic distribution of spin axes in space, and to each of 720 test particles we randomly assign a value from the interval $\pm(da/dt)_{max}$, with $(da/dt)_{max}$ being the maximum of the semi-major axis drift speed. For maximum Yarkovsky induced drift in semi-major axis, we adopted the value of 8$\times$10$^{-4}$~au~Myr$^{-1}$ obtained in Section~\ref{ss:age}. As the Yarkovsky effect scales as $da/dt \propto 1/D$, the asteroids' diameters are used to calculate the corresponding maximum value of $da/dt$ for each object, by scaling from the reference value derived for a 1~km asteroid.

While the isotropic spin axis distribution is expected after the break-up event, a post-impact spin axis evolution due to the YORP effect or non-destructive collisions may alter this assumption. Therefore, without this evolution included in the model, the assumption of the isotropic spin axis distribution has some limitations. Still, concerning simulations of the long-term dynamical evolution of asteroid families, constant Yarkovsky induced mobility in the semi-major axis of family members may be considered as the average drift rates over a postulated time-scale. Therefore, a statistically significant sample of test particles should mimic the global evolution of the family realistically, except maybe in case of some specific family substructures, such might be a \emph{YORP eye} investigated by \citet{2019MNRAS.484.1815P}. 

Finally, to have the same number of particles moving inwards and outwards, for each of 720 test particles, we created another 720 test particles with the same orbits, and with the same nominal $da/dt$, but with opposite $da/dt$ sign. In this way, we produced a total of 1440 test particles used to simulate the dynamical evolution of the Karma asteroid family.

\subsection{Dynamical evolution}
\label{ss:dyn}

\begin{figure*}
	\includegraphics[width=\textwidth]{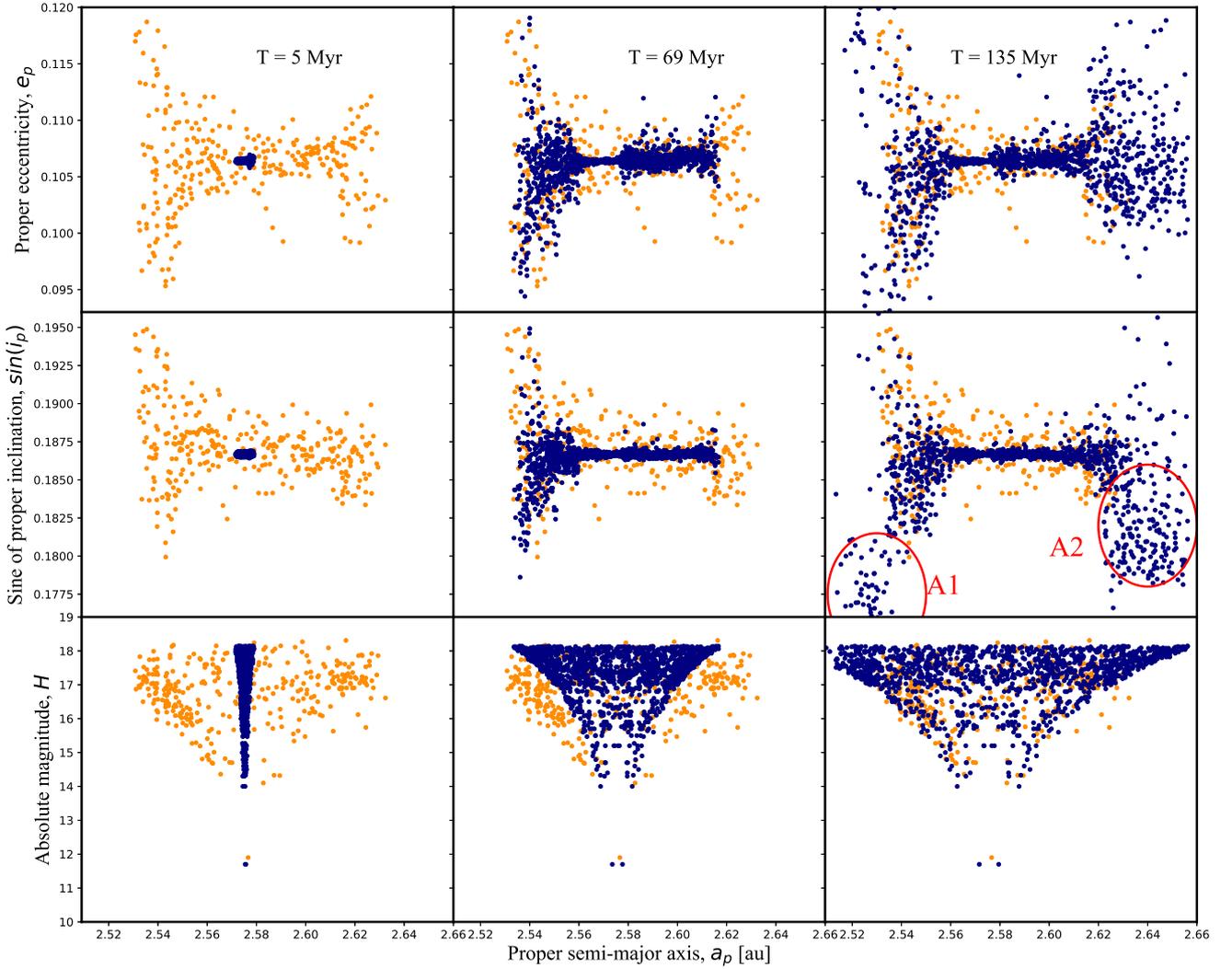}
    \caption{Snapshots from the numerical simulations of the dynamical evolution of the Karma family. Light-orange points mark the real family members, while dark-blue points denote the test particles. Two allocated regions (A1 and A2), that are filled by the test particles but no real member is located in these areas, are circled in red.}
    \label{fig:evolution}
\end{figure*}

The results of the simulations of the Karma family dynamical evolution could be summarised in the following three main phases: (i) An initially compact family begins to spread in terms of orbital semi-major axis due to the Yarkovsky effect. Drifting away from the centre of the family, particles are encountering different week mean-motion resonances. While the Yarkovsky effect is successfully moving the objects across these week MMRs, they are dispersed a bit in terms of orbital inclination, and even more in terms of orbital eccentricity.
(ii) Once the particles moving inwards crossed the weak MMRs, at the inner edge of the family they start encountering the powerful 3J:1A resonance. The latter significantly additionally disperses the eccentricities and inclinations of the encountering family members and even removes some of them from the family.
(iii) The particles moving outwards, encounter the $s-s_6-g_5+g_6$ secular resonance near the outer edge of the family. This secular resonance has a similar but less pronounced effect on the orbits of the Karma family members, as the 3J:1A resonance has at the inner edge of the family. These mechanisms produce the orbital distribution of the Karma family members that are significantly more dispersed near the family edges, fully in agreement with the observed distribution, as can be seen in Figs.~\ref{fig:aei} and \ref{fig:evolution}.

Therefore, the simulations of the dynamical evolution of the Karma asteroid family reproduce the shape that largely satisfies the shape of the family given by the observations. However, even if the family is reproduced fairly well, some regions have test particles but not real family asteroids. 
Let us note here that, as mentioned above, we use the barycenter of the family as a centre of our synthetic family.This leads to slightly shifted centre towards smaller semi-major axis concerning the location of the largest fragment, asteroid (3811)~Karma. However, the synthetic particles reach the inner edge of the family a bit earlier than the outer age (Fig.~\ref{fig:evolution}). Therefore, instead of the barycenter, a better choice for the centre of the synthetic family would be the current location of the asteroid (3811)~Karma. However, as this has only a minor effect on our results while correcting it would require redoing computationally expensive numerical simulations, we opted to keep our initial choice for the centre of the synthetic family.

Further analysis of the distribution of test particles in the $a_p - \sin(i_p)$ and $a_p - e_p$ planes, displayed in Fig.~\ref{fig:evolution}, shows that after 135~Myr of the evolution, there are some allocated parts in the test family, not visible in the real family. These two parts are circled in red and denoted as A1 and A2. Both allocated parts are located on outskirts of the family and therefore spread over the ranges of the semi-major axis not covered by the real family members. Under these circumstances, the evolution in the $a-H$ plane (see the third row in Fig.~\ref{fig:evolution}) tells of which magnitudes are test particles located in the A1 and A2 regions. We found that allocated parts consist of asteroids with magnitudes larger than $H$ = 17, well beyond the completeness limit for the  Karma family. Given that, we believe that real family members are present in the A1 and A2 regions, but most of these objects are yet to be discovered.  

To test our hypothesis, we determine how many real family members are expected in the regions where currently identified members are missing. This could be obtained directly from the simulation, and the result shows that there should be about 35 and 67 family members up to $H$ = 18.1 magnitudes, in A1 and A2 regions respectively. Then, we estimated what fraction are the family members in the whole asteroid populations. Interestingly, we found that in A1 region expected fraction of family members in the whole population of asteroids is below 30 per cent, while in the regions inhabited by the asteroids successfully associated to the family, this fraction is always above 50 per cent, and occasionally reaches even 90 per cent. Combined with a large number of still undetected main-belt asteroids in a range of 17$-$18.1 of absolute magnitudes, the number density of family members in these regions is relatively low, and this may explain why there are yet no real asteroids linked to the Karma family in the A1 and A2 allocated regions, that are filled by the test particles in our simulations.

\begin{figure}
\includegraphics[width=0.95\textwidth,angle=-90]{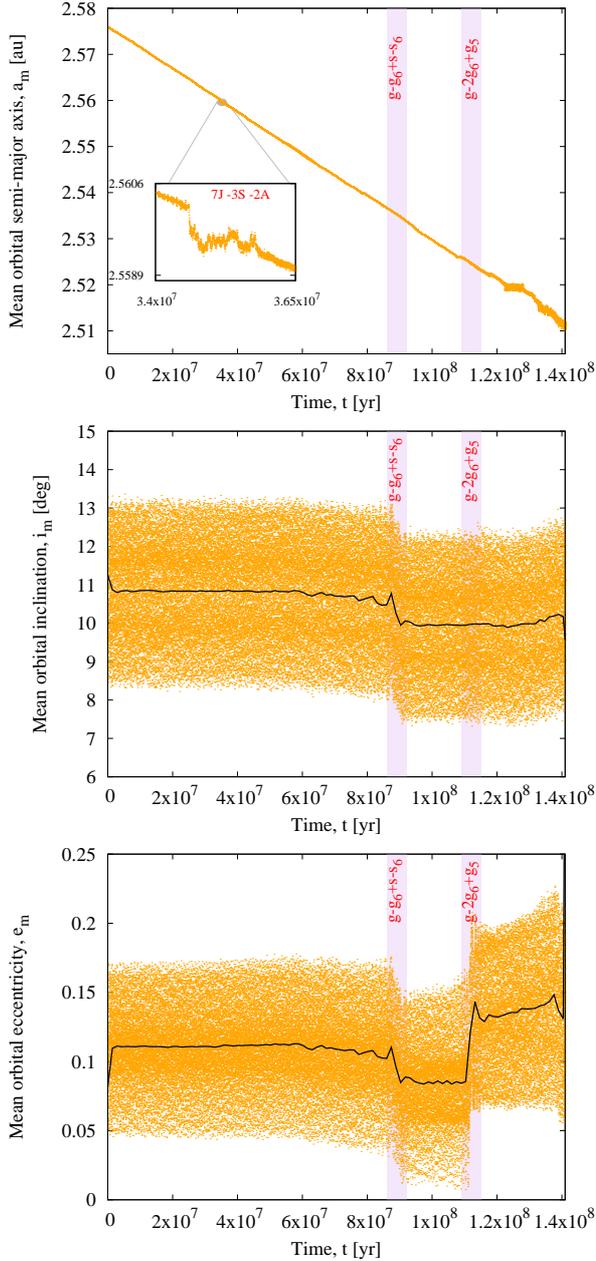}
\caption{Illustration of a dynamical route to the A1 region. The plot shows the evolution of the mean orbital semi-major axis (upper panel), inclination (middle panel) and eccentricity (bottom panel) of a test particle towards the A1 region. Scattered points represent the output from the simulations, while solid lines are obtained as the Bezier fits to the data. Vertical shaded areas, in all panels, mark periods when the particle interacts with the secular resonances indicated in the plot. Zoomed area in the upper panel shows evolution around 35~Myr when the drift in the semi-major axis is temporarily interrupted by the 7J -3S -2A mean-motion resonance.}
\label{fig:a1_tp}
\end{figure}

From the dynamical point of view, it also interesting to examine how exactly the A1 zone has been formed. To address this question, we identify in our simulations a test particle ending up in this region and analysed its dynamics. In Fig.~\ref{fig:a1_tp} we show typical orbit evolution of a test particle towards the A1 region. The critical role in this process is played by  $g-g_{6}+s-s_{6}$ secular resonance, located at a semi-major axis of about 2.535~au, for typical orbital eccentricity and inclination of the family members (see also Fig.~\ref{fig:aei}). As can be seen in Fig.~\ref{fig:a1_tp}, the orbital semi-major axis decreases due to the Yarkovsky effect, while the orbital eccentricity and inclination remain almost constant, until the test particle reaches the $g-g_{6}+s-s_{6}$ resonance. Being this a non-linear secular resonance that includes both nodal and perihelion frequencies, it affects both, orbital eccentricity and inclination. These orbital parameters are both decreased due to the resonance passage. Later on, the particle is also crossing a $g-2g_{6}+g_{5}$ secular resonance. However, as the latter resonance involves only perihelion frequencies, it does not affect the orbital inclination, but the only eccentricity, which is increased in the interaction with the $g-2g_{6}+g_{5}$ resonance. Finally, the particle is approaching the 3J:1A mean-motion resonance and its eccentricity is additionally randomly scattered. In this way, the only remaining pattern in the affected objects is their comparatively low inclinations.

\begin{figure}
\includegraphics[width=0.7\columnwidth,angle=-90]{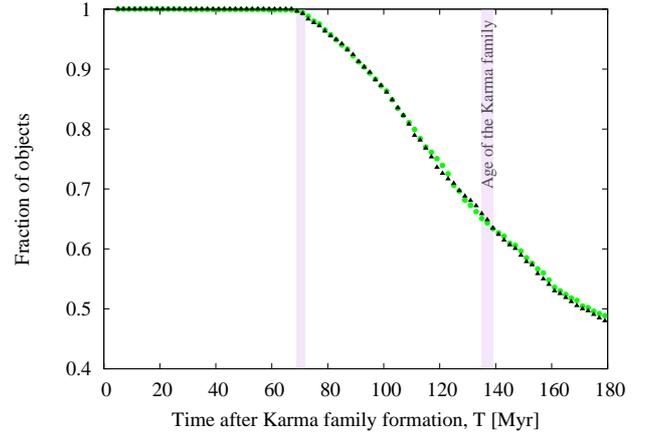}
\caption{Evolution of the number of test particles in a box defined in
3-D orbital space ($a_p$,$e_p$,$\sin$($i_p$)) by the spread of the real family members. The green points represent simulations with an extended dynamical model that also includes asteroids Ceres and Vesta, while the black triangles are results from the nominal model that includes seven major planets, from Venus to Neptune.}
\label{fig:nt}
\end{figure}

We also investigated the escape rate from the family in the course of our simulations. Fig.~\ref{fig:nt} shows the number of test particles located in the region inhabited by the known family members, as a function of time. The first particles escape the box after about 70~Myr of the evolution, and the number of particles located inside the box drops to 482 after 137~Myr (about 65 per cent of the initial synthetic population). This demonstrates that the Karma family over its estimated lifetime lost about 1/3 of the members with $H<$18.1~mag.

\begin{figure}
\includegraphics[width=0.7\columnwidth,angle=-90]{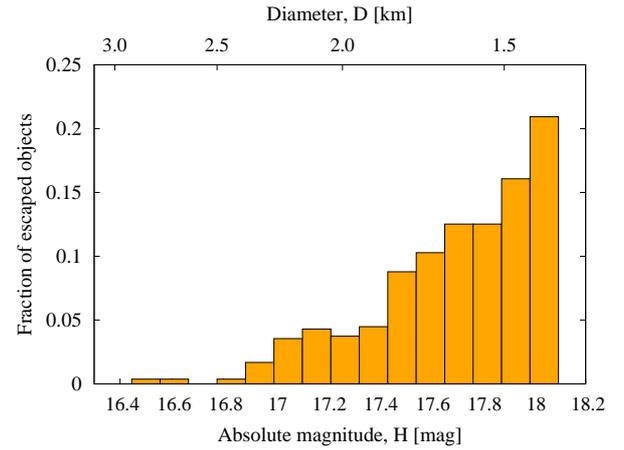}
    \caption{Magnitude distribution of the test particles that escaped during the simulation from the region of the Karma family over its lifetime. In total, 238 out of 720 particles escaped the area.}
    \label{fig:escaped_mag}
\end{figure}

The magnitude distribution of escaped particles is shown in Fig.~\ref{fig:escaped_mag}. The results reveal that all but 4 of these objects are brighter than H=$17$~mag, with the brightest escaped particle being of $H$=16.5~mag, with the latter magnitude approximately corresponding to a diameter of 2.84~km. When combined with the estimated observational completeness limit for the Karma family asteroids that is around $H$=16~mag, our findings imply that family membership up to this magnitude limit, i.e down to 3.6~km in size, is not affected by the dynamical evolution nor by the observational incompleteness. After removing interlopers from the family (see Section~\ref{ss:members}), the estimated slope of the cumulative magnitude-frequency distribution for $H\in[15,16]$~mag is $\gamma$=0.55. Therefore, this slope should be a direct result of a collisional grinding of an original population created in the disruption event that produced the family. The fact that collisional lifetime of an asteroid of 3.6~km in diameter is several times longer than the age of the family \citep{bottke2005}, indicates that the current magnitude-frequency distribution of the family members is close to the original one.

\subsubsection*{The role of Ceres and Vesta}

It is known that two largest objects in the asteroid belt, namely (1)~Ceres and (4)~Vesta may significantly perturb orbital motion of asteroids. The perturbing mechanisms include mean-motion resonances \citep{christou2012}, close encounters \citep{carruba2003} or secular resonances \citep{nov2015ApJ,2016IAUS..318...46N}. In Section~\ref{ss:dyn_env}, we concluded that the secular resonance $g-g_v+s-s_v$ with asteroid Vesta is crossing the family. Therefore, we made some additional simulations with the Ceres and Vesta added to our main dynamical model described above, in order to access their possible role in dynamics of the Karma family asteroids.

In this respect, we primarily analysed the escape rate from the family in the dynamical models with and without Ceres and Vesta included. These results are shown in Fig.~\ref{fig:escaped_mag}, and they reveal that the rate of the family members erosion is practically the same in the both dynamical models. For this reason, we concluded that the two most massive asteroids do not play an important role in dynamics of asteroids belonging to the Karma family.

\subsubsection{Flux towards the near-Earth-object region}

Being located just at the outer edge of the 3J:1A mean motion resonance with Jupiter, the Karma family is expected to provide some of its members to this resonance. Most of the asteroids that enter this resonance are later on transported towards the near-Earth-object (NEO) region. 
We already noted that our simulation consists of test particles up to $H$=18.1~mag and that some of them escape from the family via the 3J:1A route. Here we further constrain the flux from the family towards the NEO region, by extending our analysis up to $H$=18.75~mag, which is equivalent to the diameter of about 1~km, with an average albedo of Karma members.

For this purpose, following the same approach as described in Section~\ref{ss:initial_cond}, we generated 1575 test particles that represent the synthetic Karma family down to $1$~km in diameter. Each particle was given both positive and negative Yarkovsky induced drift, which resulted in a total of 3150 test particles. Following numerical simulation of the family evolution with this number of bodies, we counted how many of them entered the 3J:1A resonance. In this case, we adopted an outer edge of the resonance in terms of the semi-major axis to be 2.515~au.

\begin{figure}
\includegraphics[width=0.72\columnwidth, angle = -90]{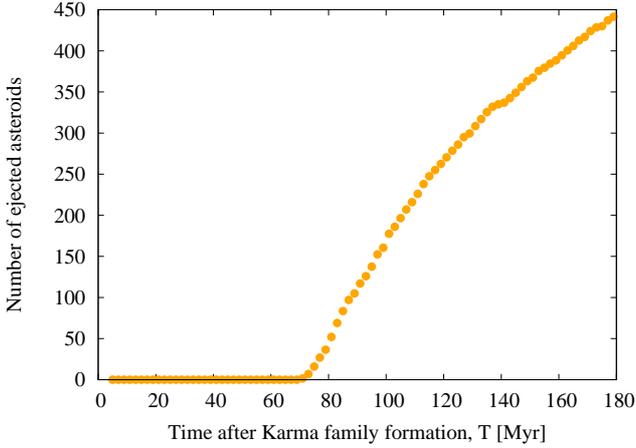}
    \caption{Number of the Karma family members, larger than 1~km in diameter, entering the 3J:1A mean motion resonance with Jupiter.}
    \label{fig:escaped_3_1}
\end{figure}

The number of family members entering the resonance as a function of time is shown in Fig.~\ref{fig:escaped_3_1}. The result shows that since its formation, the Karma family delivered in total about 350 members larger than 1~km to the 3J:1A resonance. The escape rate is almost constant, and it is about 5 objects larger than 1~km in diameter per million years.
Estimated residence time (lifetime) for main-belt asteroids arriving in the NEO region through 3J:1A resonance is about 2~Myr \citep{bottke2002, granvik2018}, which implies that there are 10 family members above 1~km in diameter present in the NEO region. Certainly, a number of sub-kilometre objects should be much larger.

This is likely a lower limit number of family members entering the 3J:1A resonance because YORP should have enough time to drive the spin states of many objects towards the extreme values, causing these members to drift towards the outskirts of the family even faster than we assumed in our simulations. Therefore, the Karma family is a non-negligible source of primitive near-Earth asteroids, supplying at least 6-7 per cent of C-complex objects above 1~km in size. 

\subsection{Constraints on the original ejection velocity field}
\label{sec: Vej_constr}

Performed numerical simulations indicate that the dynamical evolution of orbital inclinations of the Karma family members, in the semi-major axis range from 2.56 to 2.61~au, is practically negligible. Therefore,  the inclinations should be useful to infer some direct information about the initial velocity field. Surprisingly, the considered part of the family has a width of about $5\times10^{-3}$ in terms of the sine of proper inclination. This is about 25 per cent larger spread than in case of the Hoffmeister family\footnote{Note also that both families are of the same taxonomic type.} \citep{carruba2017}, even though the parent body of the latter family was at least twice as big as the parent body of the Karma family.

To further investigate this issue, we can use the proper $(a_p - \sin(i_p))$ distribution of family members in a range in proper $a_p$ between 2.56 and 2.61~au, to set constraints on the ejection velocity parameter $V_{EJ}$. Assuming that, in first approximation, the original ejection velocity field of the family could be approximated as isotropic, following the methods introduced in \citep{vok2006, Vokrouhlicky_2006b, Vokrouhlicky_2006c}, and also used in \citet{Carruba_2016}, we can simulate the ejection velocities distribution of asteroids with a Gaussian distribution, with a standard deviation that is given by:
\begin{equation}
        V_{SD}=V_{EJ}\cdot(5km/D),
    \label{eq: V_EJ}
    \end{equation}
where $D$ is the asteroid diameter and $V_{EJ}$ is the terminal ejection velocity parameter to be estimated.  Interested readers can find more details on this approach in the references above reported. The diameters of simulated objects were obtained from the absolute the magnitude of the real population of asteroids in the area, and assuming a geometric albedo of $p_v = 0.055$. 

\begin{figure*}
  \centering
    \begin{minipage}[c]{0.48\textwidth}
    \centering \includegraphics[width=3.2in]{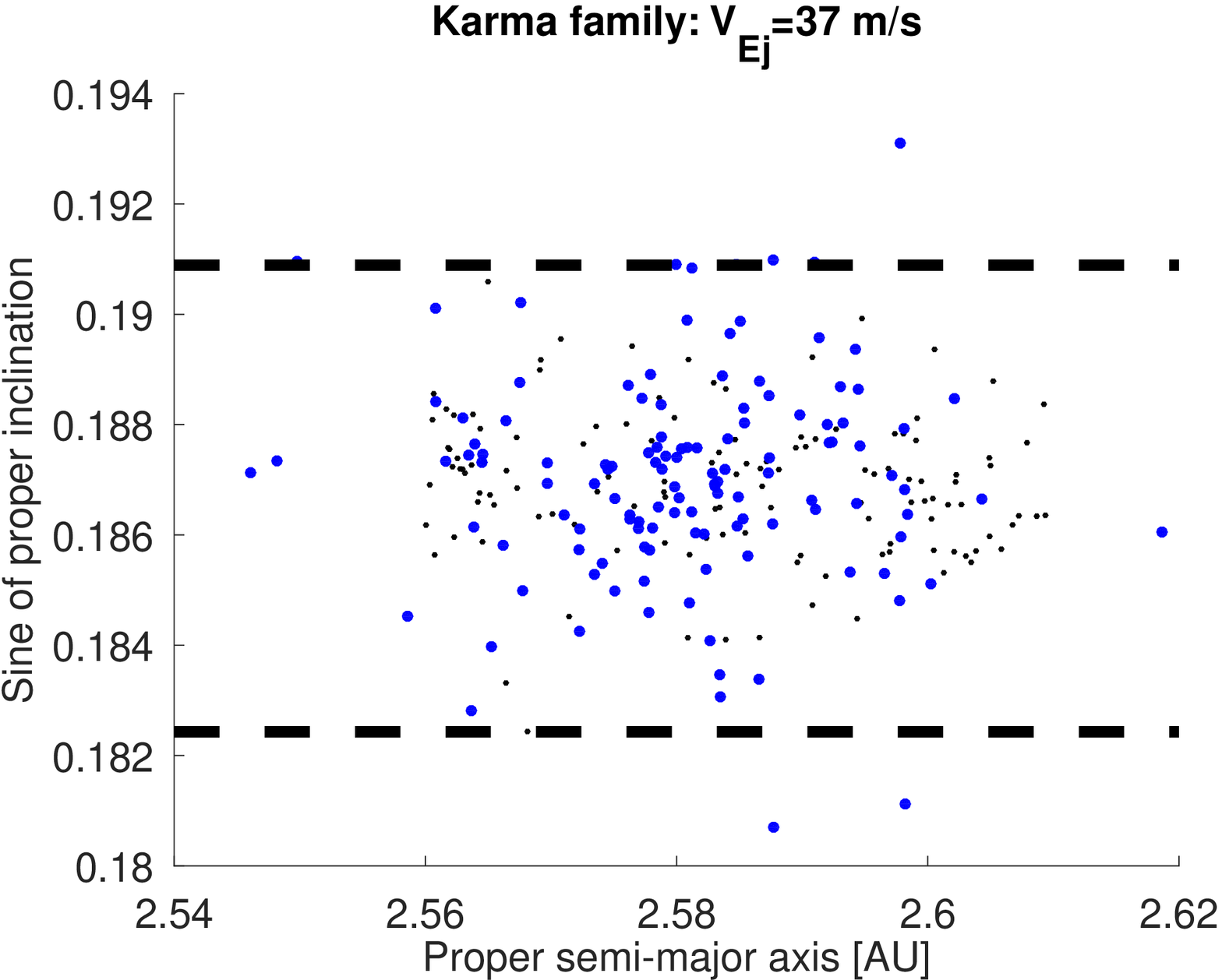}
  \end{minipage}%
  \begin{minipage}[c]{0.48\textwidth}
    \centering \includegraphics[width=3.2in]{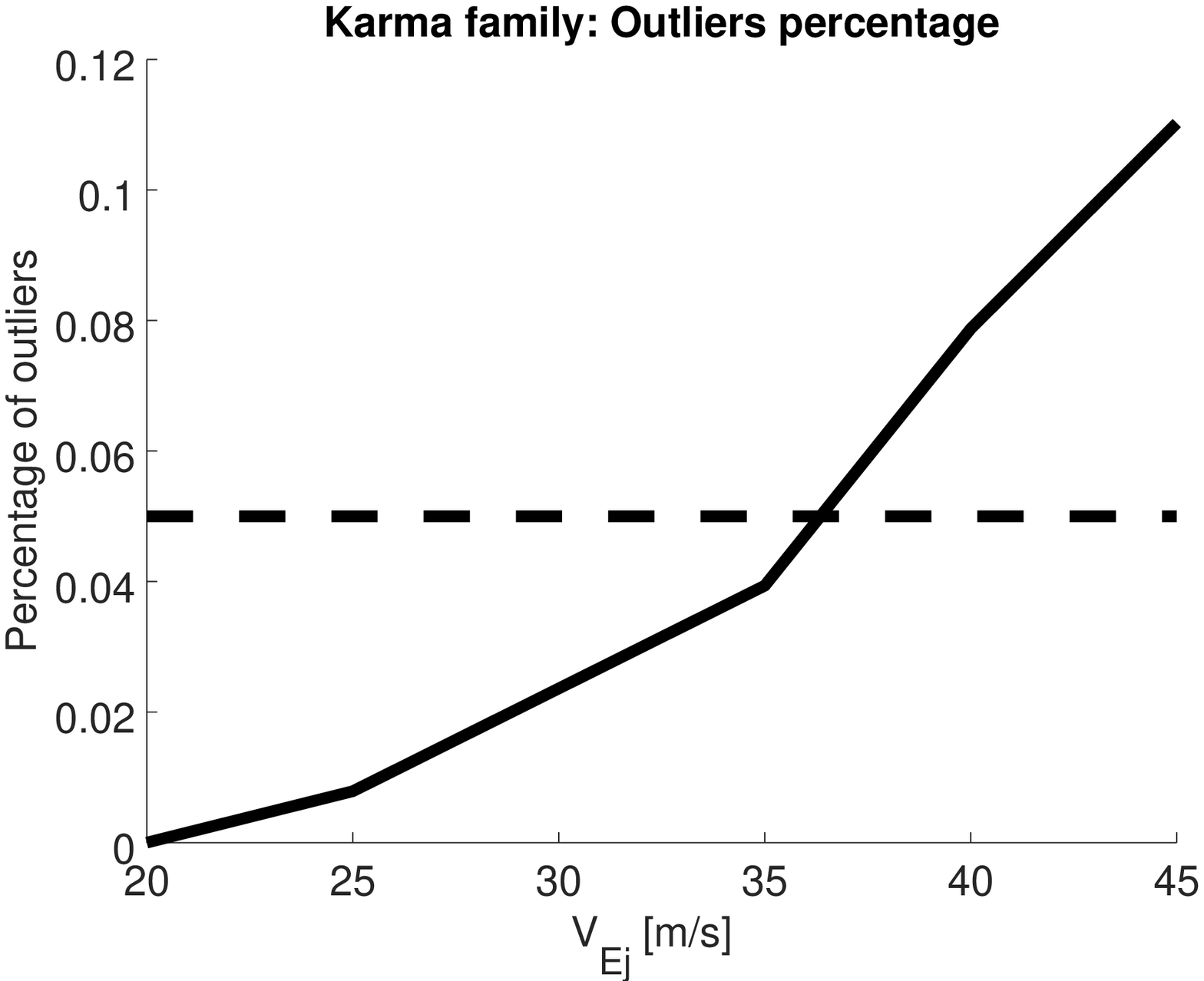}
  \end{minipage}
  
\caption {Left panel: An $(a_{p}, \sin{(i_{p})})$ plot of members of the Karma family in the range of semi-major axis from 2.56 to 2.61~au (black dots).
  Horizontal dashed lines display the minimum and maximum value of
  $\sin{(i_p)}$ for these objects.  Blue full circles show the orbital location of simulated family members for a $V_{EJ}$ parameter equal to 37~m$s^{-1}$. Right panel: the percentage of simulated asteroids that
  lies out of the area between dashed lines, as a function of the $V_{EJ}$
  parameter.  The horizontal dashed line displays the 5 per cent probability level.}
\label{Fig: Karma_vej}
\end{figure*}

We simulated the asteroid family for various values of the $V_{EJ}$ parameter and counted the number of simulated members with values of $\sin{(i)}$ outside the range of those observed for real family members.  We call test particles outside this range outliers. The left panel of Fig.~\ref{Fig: Karma_vej} displays the $a_p - \sin(i_p)$ distribution of real and simulated family members for the optimal value of the $V_{EJ}$ parameter, while the right panel shows the percentage of outliers as a function of $V_{EJ}$. For $V_{EJ}$ = 37~m$^{-1}$ the percentage of generated outliers is higher than 5 per cent, which is generally associated with the null hypothesis.  This sets an upper limit on the value of the $V_{EJ}$ parameter of the original ejection velocity field. Since the estimated escape velocity ($V_{esc}$) from the parent body of the Karma family should not exceed 17~m$s^{-1}$ (see Section~\ref{ss:initial_cond}), this yields a value of $\beta = V_{EJ} / V_{esc} = 2.2$, which is somewhat larger than the values observed for most asteroid families, which have a $\beta$ in the range 0.5 to 1.5 \citep{Carruba_2016}. Even if we assume that the parent body was significantly larger, the normal component of the ejection velocity field seems too high. For instance, assuming that the parent body was 50~km in diameter, which should be an upper limit, the resulting escape velocity would be only about 21~m$s^{-1}$. This would yield the corresponding parameter $\beta$ of ~1.8, still slightly outside the range of its typical values.

There are at least three possible explanations for the large value of the $\beta$ parameter: (i) it may be due to an anisotropic initial velocity field, characterised by an unusually large velocity component normal to the orbital plane, as is the case for instance with the Veritas asteroid family \citep{novetal2010}, (ii) alternatively, it may be because of unusually large impact velocity, or (iii) it might be possible that some asteroids associated to the family, in proper $a_p$ between 2.56 and 2.61~au, are not real family members, but linked to the family due to the HCM chaining effect.

\section{Summary and Conclusions}

In this paper, we investigated different characteristics of the Karma
asteroid family. Our main results are summarised below.

\begin{itemize}
  \item The number of asteroids associated to the family by the HCM, at velocity cut-off of $v_c=55$~m$s^{-1}$, is 332,and these are dark asteroids with an average geometric albedo of $p_v=0.055$. We excluded 15 objects with $p_v>0.1$,
  which are likely interlopers, yielding to 317 currently known Karma family members.
  \item The family is formed in a catastrophic disruption of a parent body that was likely between 34 and 41~km in diameter. Still, due to a possible contribution of yet undetected asteroids, the real size of the parent body could be larger.
  \item  Based on the slopes of the family V-shape in the ($a_p$ - $1/D$) plane and under the certain assumptions described in Section~\ref{ss:age}, the age of the Karma family is $137\pm21$~Myr. 
  \item Using the numerical simulations of the long-term dynamical evolution of the family, we successfully reproduced the overall shape of the family but also identified some parts of the family still not recognised by the HCM identification approach.
  \item Our numerical simulations also indicate that some of the Karma family members have been reaching the near-Earth region, via the 3J:1A mean-motion resonance. About 10 asteroids larger than $1$~kilometre, originating in the family, are expected to reside currently in the region of terrestrial planets. 
  \item Characterisation of the ejection velocity field, based on the component related to the orbital inclination, indicates approximately factor 2 larger ejection velocities than typically found in asteroid families.
  
\end{itemize}

\section{Future Prospects}

The origin of the Karma family parent body may be an interesting subject to investigate. Our analysis of the Karma ejection velocities points out different behaviour with respect to other primitive families, such as the Hoffmeister family. Although both families are of primitive composition, the parent body of the Hoffmeister family was about twice as large as the parent body of the Karma family, while the ejection velocities have on average a factor of about 2 larger in the case of the latter family. Porous targets are found to have a significantly higher impact strength than the rubble-pile parent bodies and show a behaviour more similar to non-porous monolithic targets \citep{2019Icar..317..215J}. Therefore, different ejection velocities in case of the two families may be diagnostic of porosity of their parent bodies. In this connection, especially helpful would be the numerical simulation of the impact disruption and re-accumulation \citep{2012Icar..219...57B,2019Icar..317..215J} of the Karma family parent body.

The origin of primitive parent bodies is also very important for understanding the origin and amount of the water-ice content in asteroids. A population of known main-belt comets (MBCs), objects whose observed activity is believed to be driven by the sublimation of volatile ices is constantly growing \citep[e.g.][]{Snodgrass2017}. Still, several questions regarding their structure, composition, and origin are yet to be answered. For instance, the MBCs are found to be associated with collisional asteroid families of primitive taxonomic classifications \citep{Hsieh2018,nov2018}. The low thermal conductivity of porous cometary material suggests that ice may be retained in the interior of main-belt asteroids, despite continual solar heating. Thermal models have shown that it may be possible for water-ice to be preserved on main-belt asteroids over the age of the Solar System \citep{2008ApJ...682..697S}. Moreover, under some circumstances, such as continuously small rotational axis tilt, some asteroids may preserve the ice even in the NEO region \citep{2020Icar..34813865S}. Still, all the known MBCs have their semi-major axis beyond 2.7~au. Observational studies of asteroids from the Karma family may provide a new view on the inner edge of icy-asteroids in the main belt, and also possible in the NEO region. In this respect, targeting observations of family members in search for signs of activity, for instance, those performed by \citet{nov2014} among members of Gibbs asteroid cluster may be very useful. Also, for discovering possible water-features in spectra of the Karma family asteroids, particularly useful could be space-based observations such are for instance those planned within \textit{Twinkle} mission \citep{twinkle2019}.

According to our results, the first members of the Karma family arrived in the NEO region about 65 Myr ago. At this time the Chicxulub impactor struck the Earth, causing the Cretaceous/Tertiary (K/T) mass extinction event. It was argued by \citet{bottke2007} that the Baptistina asteroid family is the most likely source of the K/T impactor. However, it was shown later by \citet{reddy2011} that the composition of the Baptistina family members does not match the composition of the K/T impactor that likely was a CM2-type carbonaceous chondrite \citep[e.g.][]{1998Natur.396..237K,2006E&PSL.241..780T}. Therefore, these findings make the Baptistina family an unlikely source of the K/T impactor.
The estimated flux of the Karma family members towards the Near-Earth region, and even more their sizes, are both too small to suggest this family as a source of the K/T impactor. Still, any study aiming to investigate the impact rate on terrestrial planets in this period should take into account a contribution of the Karma family.

Even though the Karma family may not be a good candidate to be linked with the K/T impactor, about 350 of its members entered the NEO region through the 3J:1A resonance. According to \citet{granvik2018} and \citet{bottke2006}, probability of striking the Earth for the main belt asteroid escaping by this route is between 0.1 and 0.3 per cent. Therefore, a probability that one of the escaped Karma family members larger than 1~km in diameter impacted our planet is about 50 per cent. This implies that a km-sized family member possibly hit the Earth in the last 70~Myr. Such impactor would form a crater of about 20~km in diameter. Apropos, it would be very interesting to see if any of the Earth's craters of the given age and size, may be linked to the impact of a carbonaceous chondrite body.

\section*{Acknowledgments}
We would like to thank David Vokrouhlick{\'y} for his careful review, and comments which we found helpful in improving the quality of the manuscript. BN and VR acknowledge support by the Ministry of Education, Science and Technological Development of the Republic of Serbia, contract
No. 451-03-68/2020-14/200104.  VC is grateful to the S\~{a}o Paulo State Science Foundation (FAPESP, grant 2018/20999-6)  and to the Brazilian National Research Council (CNPq, grant 301577/2017-0). 

\section*{Data availability}

The data underlying this article will be shared on reasonable request to the corresponding author.







\bsp	
\label{lastpage}
\end{document}